\documentstyle[prd,aps,floats,epsfig]{revtex}
\tighten

\newcommand{\al}{\alpha}

\newcommand{\ga}{\gamma}

\newcommand{\la}{\lambda}

\newcommand{\th}{\theta}

\newcommand{\ra}{\rightarrow}

\newcommand{\be}{\begin{equation}}
\newcommand{\ee}{\end{equation}}

\newcommand{\bea}{\begin{eqnarray}}
\newcommand{\eea}{\end{eqnarray}}
\newcommand{\bean}{\begin{eqnarray*}}
\newcommand{\eean}{\end{eqnarray*}}
\newcommand{\dd}{\partial}

\newcommand{\bb}{\bibitem}     
\def\spose#1{\hbox to 0pt{#1\hss}} 
\def\ltapprox{\mathrel{\spose{\lower 3pt\hbox{$\mathchar"218$}} 
 \raise 2.0pt\hbox{$\mathchar"13C$}}} 
\def\gtapprox{\mathrel{\spose{\lower 3pt\hbox{$\mathchar"218$}} 
 \raise 2.0pt\hbox{$\mathchar"13E$}}} 
\def\inapprox{\mathrel{\spose{\lower 3pt\hbox{$\mathchar"218$}} 
 \raise 2.0pt\hbox{$\mathchar"232$}}} 
 
\begin{document}
\draft
\preprint{\ 
\begin{flushright}
%CERN-TH/99-xxx \\
UGVA-DPT 1999/12-zzzz\\
astro-ph/9912081 \\
\end{flushright}
}

\twocolumn[\hsize\textwidth\columnwidth\hsize\csname@twocolumnfalse\endcsname 
\title{Anisotropic 'hairs' in string cosmology}
\author{ Kerstin E. Kunze and Ruth Durrer}
\address{ D\'epartement de Physique Th\'eorique, Universit\'e de Gen\`eve,
24 quai Ernest Ansermet, CH-1211 Gen\`eve 4, Switzerland}

\maketitle

\begin{abstract}
In this letter we investigate whether the isotropy problem is
naturally solved in inflationary cosmologies inspired by string theory,
so called pre-big-bang cosmologies. We find that, in contrast to what
happens in the more common 'potential inflation' models, initial
anisotropies do not decay during pre-big-bang inflation.
\end{abstract}

\date{\today}

 ] 

In most models, inflation is driven by the potential energy of a scalar
field. This has similar effects as adding vacuum energy or, equivalently, a
positive cosmological constant. For these models Wald \cite{wald}
proved a cosmic no hair theorem, which implies that an initial shear
is inflated away \cite{tw}.

A couple of years ago, a new mechanism of inflation has been
proposed, where inflation is due to the kinetic term of the dilaton
which is always present in the low energy effective action from a
string theory. These string cosmologies are symmetric under the duality
transformation,  $t\ra -t$ and $a\ra 1/a$. Here $t$ is cosmic time and
$a$ is the scale factor. An expanding solution at negative times is
inflating~\cite{PBB}. Here, we want to study whether a 'no hair'
theorem is also valid for these pre-big-bang inflationary solutions.

We investigate especially spatially homogeneous, but anisotropic cosmologies
in the pre-big-bang scenario. The aim is to determine the evolution of
a possible primordial shear. The pre-big-bang universe starts off at
$t\ra -\infty$ in a nearly Minkowski spacetime. But
the hypothesis of low curvature and low  coupling in the
pre-big-bang scenario does not say anything about the possibility 
of starting with an anisotropic spacetime.
 
Recently, the initial conditions of the pre-big-bang scenario have
been criticized not to be natural~\cite{crit}. 
 Buonanno et al.~ \cite{bub} have  addressed this issue and have 
shown that the pre-big-bang inflationary phase in the string frame 
is equivalent to gravitational collapse in the Einstein frame. 
They therefore have
concluded that the initial conditions for pre-big-bang inflation are
as natural as those for gravitational collapse.
However, the pre-big-bang bubble picture  suggested in this work
\cite{bub} seems to favor anisotropic Kasner solutions. 
It is thus important to investigate whether such initial anisotropies
are inflated away during the subsequent inflationary evolution.

Below we show that this is not the case. We find
that the behaviour of shear in pre-big-bang cosmology is quite  
different from its behavior in ordinary potential inflation
and primordial shear is not inflated away.

We first  briefly recall
the expressions for the Ricci tensor 
of spatially homogeneous models in the orthonormal  frame.
(Latin indices run from
0..3 and Greek indices from 1..3.)
The metric $(g_{ab})$ of spatially homogeneous models can be written
as
\begin{eqnarray}
ds^{2}=-dt^2+h_{\alpha\beta}(t)\omega^{\alpha}\omega^{\beta}
\end{eqnarray}

where $\{\omega^{\alpha}\}$ is an invariant basis of one-forms
satisfying the algebra
$$d\omega^{\alpha}=\frac{1}{2}C^{\alpha}_{\mu\nu}\omega^{\mu}
\wedge\omega^{\nu},$$
where $C^{\alpha}_{\mu\nu}$ are the  structure
constants of the symmetry group of the corresponding homogeneous model. 
Bianchi models are divided into class A and B depending on properties
of the group structure constants. Here we restrict ourselves to
Bianchi class A models which are characterized by 
$$C^{\alpha}_{\mu\alpha}=0.$$ For these models $h_{\alpha\beta}(t)$ is
diagonal and thus of the form
\be
h_{\alpha\beta}(t)=diag\left(a_{1}^{2}(t),a_{2}^{2}(t),
a_{3}^{2}(t)\right).
\ee

We also choose an orthonormal
frame $\sigma^{(a)}$, so that 
\be
ds^2=\eta_{(a)(b)}\sigma^{(a)}\sigma^{(b)}~,
\ee
where $\eta_{(a)(b)}=diag(-1,+1,+1,+1)$.
Indices in parentheses refer to the orthonormal frame.

The relation between the basis one-forms and the orthonormal frame is given by,
$\sigma^{(\alpha)}=a_{\alpha}(t)\omega^{\alpha}$ (no sum over $\alpha$).

The parameter $t$ is chosen such that $g^{ab}n_{a}n_{b}=-1$, where
$n=-{\dd\over \dd t} $ is the normal to the (space-like) homogeneous
hyper-surfaces. The expansion  and shear tensors, $\theta_{ab}$
and $\sigma_{ab}$ respectively, of the hyper-surfaces $\{ t= $ const.$\}$ are 
defined by \cite{el-mac}
\begin{eqnarray}
n_{a;b}&=&\theta_{ab}\\
\sigma_{ab}&=&\theta_{ab}-\frac{1}{3}\theta(g_{ab}+n_{a}n_{b})~,
\end{eqnarray}
where $\theta=\theta^{a}_{a}$ is the expansion.
In terms of the scale-factors
$\theta=\sum_{\al=1}^{3}\frac{\dot{a}_{\al}}{a_{\al}}$. 
A dot indicates derivative with respect to $t$.
The shear $\sigma_{ab}$ is trace-free.  It is convenient to define
\be
\sigma^{2}\equiv\frac{1}{2}\sigma_{ab}\sigma^{ab}.
\ee
With this notation the non-vanishing 
components of the Ricci tensor for Bianchi class A 
models in the 
orthonormal frame are given by (for useful formulae see \cite{chandra}
\cite{ryan})

\begin{eqnarray}
R_{(0)(0)}&=&-\dot{\theta}-\frac{1}{3}\theta^{2}-2\sigma^{2}\nonumber\\
R_{(\alpha)(\alpha)}&=&
\dot{\sigma}_{(\alpha)(\alpha)}+\theta\sigma_{(\alpha)(\alpha)}+
\frac{1}{3}\dot{\theta}+\frac{1}{3}\theta^{2}+F_{(\alpha)(\alpha)} 
	\label{Ricci}
\end{eqnarray}

where there is no sum over $\alpha$ and $F_{(\alpha)(\alpha)}$ is a 
function of the scale factors defined by
\bean
F_{(\alpha) (\alpha)} &=&
  \gamma_{(\alpha)}^{\;\;\;\;(\mu)(\kappa)}\gamma_{(\kappa)(\mu)
(\alpha)}-\gamma_{(\alpha)}^{\;\;\;\;(\mu)(\kappa)}
\gamma_{(\alpha)(\kappa)(\mu)} \\ && -\gamma_{(\kappa)(\mu)(\alpha)}
\gamma_{(\alpha)}^{\;\;\;\;(\kappa)(\mu)}~.
\eean
  The Ricci rotation coefficients, $\gamma$ are given by
\bean
\gamma^{(\alpha)}_{\;\;\;\;(\beta)(\mu)}
 &=&\frac{1}{2}
\left[
\frac{a_{\alpha}}{a_{\beta}a_{\mu}}C^{\alpha}_{\beta\mu}
-\frac{a_{\mu}}{a_{\alpha}a_{\beta}}C^{\mu}_{\alpha\beta}
+\frac{a_{\beta}}{a_{\mu}a_{\alpha}}C^{\beta}_{\mu\alpha}
\right] \\
 &=& \ga_{(\alpha)(\beta)(\mu)} =  \ga_{(\alpha)}^{~~(\beta)(\mu)} =
	\dots~.
\eean
The group structure constants $C^{\alpha}_{\beta\mu}$ for
the different Bianchi models can be found, for example, in \cite{ryan}.

For Bianchi class B models the Ricci tensor has off-diagonal
components and additional terms in the diagonal components
(see for example \cite{biaB}).
However, all these  scale with the scale factors in such a way that
they  become sub-dominant during inflationary expansion.
Therefore, the discussion of Bianchi class
A models presented here is sufficient.

Let us now derive the equations of motion for pre-big-bang inflation
in this background.
The low energy effective action of string theory is given by
\begin{eqnarray}
S=-\frac{1}{16\pi G}\int d^{4}x \sqrt{-g}e^{-\phi}(R+\partial _{\mu}\phi
\partial^{\mu}\phi) + S_{\rm matter}.
\label{act}
\end{eqnarray}
As matter source we include a perfect fluid. The equations of motion 
derived from (\ref{act}) are then given by~\cite{gab}
\begin{eqnarray}
R_{\mu}^{\nu}+\nabla_{\mu}\nabla^{\nu}\phi&=&8\pi G e^{\phi} T_{\mu}^{\nu}\\
R-(\nabla_{\mu}\phi)^{2}+2\nabla_{\mu}\nabla^{\mu}\phi&=&0\\
\dot{\rho}+\theta(\rho+p)&=&0.
\end{eqnarray}

The last equation already uses the form of the energy momentum tensor
$T_{ab}=\rho(t) n_{a}n_{b}+ p(t)(g_{ab}+n_{a}n_{b})$. We assume that the 
perfect fluid satisfies the equation of state $p=\gamma\rho$.

In order to discuss Bianchi class A space-times we use the Ricci
tensor given in Eq.~(\ref{Ricci}). In the orthonormal frame defined
above, these equations  then read
\be
\dot{\sigma}_{(\mu)(\mu)}+(\theta-\dot{\phi})\sigma_{(\mu)(\mu)}=
-F_{(\mu)(\mu)}
+\frac{1}{3}\sum_{\alpha}F_{(\alpha)(\alpha)} \label{sigma1} 
\ee
\begin{eqnarray}
\dot{\theta}+(\theta\! - \!\dot{\phi})\theta&=&
-\sum_{\mu}F_{(\mu)(\mu)}
+8\pi G e^{\phi}3\gamma\rho\label{theta1}\\
\ddot{\phi}+(\theta \! -\! \dot{\phi})\dot{\phi}&=&8\pi G e^{\phi}
(3\gamma-1)\rho\label{phi1}\\
\dot{\rho}+\theta(\gamma\! + \! 1)\rho&=&0\\
\frac{1}{3}\theta^{2}-\sigma^{2} -\frac{1}{2}(2\theta \!- \!\dot{\phi})
\dot{\phi} &=&
-\frac{1}{2}\sum_{\mu}F_{(\mu)(\mu)}
+8\pi G e^{\phi}\rho. \label{con1}
\end{eqnarray}

It is easy to see that for $\rho=F_{(\mu)(\mu)}=0$ this system is invariant
under the transformation
\be
 a_\al \ra 1/a_\al~,~ t\ra -t~ \mbox{ and } \dot{\phi} 
\ra 2\theta-\dot{\phi} ~,
\ee
the so called scale factor duality. Under these changes, an
expanding decelerating solution in the post-big-bang era ($t>0$)
transforms into an inflating expanding solution in the pre-big-bang
era ($t<0$)~\cite{gab}. For the following discussion about the  pre-big-bang
inflationary era, we have to keep in mind that $t$ goes from $-\infty$ to 0.

The evolution equation for the  matter energy density yields
\begin{eqnarray}
\rho=\frac{\rho_{0}}{\left(a_{1}a_{2}a_{3}\right)^{\gamma+1}}.
\end{eqnarray}

We assume that inflation has started and $\rho$ describes an 'ordinary'
fluid with $\gamma>-1$. As we shall check at the end, it is then
justified to neglect terms involving $\rho$. 
Furthermore, the terms  $F_{(\alpha)(\alpha)}$ can be neglected. Also this
hypothesis will be checked later for consistency.

With these approximations, Eqs.~(\ref{sigma1}) to (\ref{con1}) reduce to
\begin{eqnarray}
\dot{\sigma}_{(\mu)(\mu)}+(\theta-\dot{\phi})\sigma_{(\mu)(\mu)}&\simeq&0
\label{sigma}\\
\dot{\theta}+(\theta-\dot{\phi})\theta&\simeq&0\label{theta}\\
\ddot{\phi}+(\theta-\dot{\phi})\dot{\phi}&\simeq&0\label{phi}\\
\frac{1}{3}\theta^{2}-\sigma^{2}&\simeq&\frac{1}{2}(2\theta-\dot{\phi})
\dot{\phi}.\label{constra}
\end{eqnarray}
But this set of equations can be readily solved and with scale factors
of the form
$a_\al = (t/t_0)^{-\la_\al}$ which implies
\begin{eqnarray}
\theta=-\frac{\sum_{\alpha}\lambda_\alpha}
{t}~, \quad \dot{\phi}=-\frac{\sum_{\alpha}\lambda_{\alpha}+1}{t}~.
\label{e22}
\end{eqnarray}
Since $t$ is negative in our domain of interest and we want positive
scale factors, we choose also $t_0$ negative. Expansion in all
directions is then garanteed if $\la_{\al}>0$.

The evolution of $\sigma_{(\alpha)(\alpha)}$ is given by
\begin{eqnarray}
\sigma_{(\alpha)(\alpha)}=
\frac{ \frac{1}{3}\sum_{\mu}\lambda_\mu-\lambda_{\alpha}}{t}
\end{eqnarray}
The constraint equation (\ref{constra})
yields the Kasner constraint
\[\sum_{\alpha}\lambda^{2}_{\alpha}=1~. \]

We have thus found that the quantities  we are
 interested in, the relative amplitudes of  
shear, i.e. $\frac{\sigma_{(\alpha)(\alpha)}}{\theta}$ and
$\frac{\sigma^{2}}{\theta^{2}}$ remain constant.
A primordial shear is
not inflated away during pre-big-bang inflation,
\begin{eqnarray}
\frac{\sigma_{(\alpha)(\alpha)}}{\theta}
=\mbox{ const.} \quad\quad
\frac{\sigma^{2}}{\theta^{2}}=\mbox{ const.}
\end{eqnarray}
This is our main result.

It remains to check that it is justified 
to neglect terms involving $\rho e^{\phi}$ and the $F_{(\al)(\al)}$.
The first expression is given by
\begin{eqnarray}
\rho e^{\phi}\sim\left(\frac{t}{t_{0}}\right)
^{\gamma\sum_{\alpha}\lambda_{\alpha}-1}.
\end{eqnarray}

Consistency requires (cf. equations (\ref{theta1}),
(\ref{phi1})) that, with increasing time, this term becomes less and
less important if compared, for example, with $\dot{\th}$. In other words,
$\gamma\sum_{\alpha}\lambda_{\alpha}-1
>-2.$
For positive $\gamma$ this is always satisfied since $\la_\al>0$.
For $\gamma<0$ it leads to the constraint
\begin{eqnarray}
\sum_{\alpha}\lambda_{\alpha}<\frac{1}{\mid\gamma\mid}.
\label{gam}
\end{eqnarray}
But  the Kasner constraint which holds for the unperturbed solution, 
$\sum_{\alpha}\lambda_{\alpha}^{2}=1$, 
implies $\sum_{\alpha}\lambda_{\alpha} >\sum_{\alpha}\lambda_{\alpha}^{2}=1$.
Hence the inequality (\ref{gam}) is satisfied for $\gamma>-1$.

Let us now  discuss
the behaviour of the functions $F_{(\alpha)(\alpha)}$.
In particular, we want to address the question 
how the shear can be effected by  a 
contribution from the $F_{(\alpha)(\alpha)}$.
For self consistency, we just have to check that for a solution close
to Kasner, the $F_{(\alpha)(\alpha)}$'s may be neglected.
The dominant contribution to the Ricci rotation coefficients  comes from
a factor $\left(\frac{a_{\mu}}{a_{\alpha}a_{\beta}}\right)^2$
where $a_\mu$ expands fastest and $a_\alpha$ and $a_\beta$ expand slowest. 
For a Kasner solution the contribution to the Ricci rotation 
coefficients   grows like 
$\left(\frac{a_\mu}{a_\alpha a_\beta}\right)^2 \propto 
t^{2(\la_\al+\la_\beta-\la_\mu)}$. The term with
minimal $\la_\alpha+\la_\beta-\la_\mu=\la_{\min}$ grows fastest. 
But the Kasner condition readily implies that $0<\la_\al<1$ so that
$\la_{\min}>-1$. Therefore, if the
deviation from the 'Kasner solution' is small, it
decreases with time and will eventually be negligible.
If at some given time during pre-big-bang inflation, the universe is
close to a a Kasner solution, it will approach the Kasner solution
during subsequent evolution. In that sense the Kasner solutions are
(local) attractors of the Bianchi type A models with ordinary matter
content.

Note also that, if the solution is reasonably close to 
isotropic, $\la_\al \sim 1/\sqrt{3}$, $\la_{\min}$ is even positive 
and the $F_{(\al)(\al)}$ are very strongly suppressed. This means that
our argument applies and subsequent pre-big-bang evolution does not
'isotropize' the solution.

As a simple example we consider a Bianchi II string cosmology.
We neglect a possible additional contribution from matter,
$\rho=0$. The exact 7 parameter family of solutions can be found
in~\cite{bat}. The evolution of the ratios $\frac{\sigma_{(\alpha)(\alpha)}}
{\theta}$ for a particular choice of parameters is shown in Fig.~1. We
have chosen the parameters such that the solution converges to the
Kasner solution with scale factors
$a_1\propto (-t)^{-\lambda_{1}}$, $a_2\propto (-t)^{-\lambda_{2}}$, and
$a_3\propto (-t)^{-\lambda_{3}}$, where 
$\lambda_{1}=0.68$, $\lambda_{2}=0.61$, and $\lambda_{3}=0.42$.
These satisfy the string Kasner condition $\sum_{1}^{3}
\lambda_{i}^{2}=1$.
The evolution of the shear parameters $\sigma_{(\al)(\al)}$
for this particular pre-big-bang inflationary solution
is shown in Fig.~1. Clearly, $\sigma_{(\al)(\al)}/\theta$ approaches
the Kasner value,
\[\sigma_{(\al)(\al)}/\theta 
\stackrel{t\ra 0^{-}}{\longrightarrow} 
{\la_\al - {1\over
3}\sum_{\mu}\la_\mu \over \sum_{\mu}\la_\mu} ~. \]
\begin{figure}[ht]
\centerline{\epsfig{file=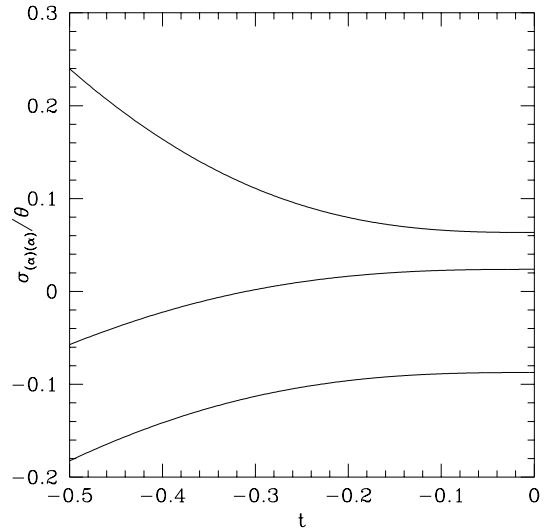, width=2.9in}}
\caption{Evolution of the ratios $\frac{\sigma_{(\alpha)(\alpha)}}
{\theta}$ in a Bianchi II model. The parameters are chosen in order
to admit an pre-big-bang inflationary solution with $\lambda_{1}\sim
0.68$, $\lambda_{2}\sim 0.61$ and $\lambda_{3}\sim 0.42$.}
\end{figure}

The situation in pre-big-bang inflation 
can be contrasted with ordinary slow roll
inflation. There the scale factor is expanding close to exponentially
and  $\th \sim$ const.
The equation for $\sigma_{(\alpha)(\alpha)}$ 
in slow roll inflation is obtained from  (\ref{sigma1})  by setting
the dilation term, $\dot{\phi}=0$.
Neglecting again the right hand side of
(\ref{sigma1}), we find in this case
 $$\frac{\sigma_{(\alpha)(\alpha)}}{\theta}
\sim e^{-\theta t}~.$$
Primordial shear decays exponentially and very soon it becomes negligible.

In contrary, in pre-big-bang inflation the shear parameter
is essentially determined by its
initial value.
This behaviour can also be understood by looking at the 
evolution in the Einstein frame in which 
usual general relativity is recovered. The dilaton 
provides a matter content which  behaves as
a stiff perfect fluid ($p=\rho$) whose
energy density evolves like $a^{-6}$. The shear 
$\sigma^{2}$ also evolves as $a^{-6}$, and again their
ratio is a constant.

It is known in vacuum general relativity that the singularity
at $t=0$ in spatially homogeneous models is either asymptotically
velocity term dominated \cite{vel} (Kasner-like)
or Mixmaster like \cite{mix}.
Adding a massless scalar field destroys the Mixmaster
behaviour after a finite number of oscillations 
leaving just the Kasner behaviour \cite{scal_mix}.
Thus at sufficiently late times, $t\ra -0$ terms due to spatial curvature
like the $F_{(\alpha)(\alpha)}$ terms in the string frame 
become negligible.
 
We consider our result as quite important for the pre-big-bang
model. It implies  that the problem of isotropization
cannot be solved in the simplest version
of pre-big-bang inflation. Especially, it cannot be solved in the
early, classical, low coupling regime. We also doubt that this problem
can be solved by quantum particle production back-reaction, a
mechanism, which can be used to some extent to damp anisotropies in the
very early post-big-bang universe~\cite{Dorshk}. In our case, the 
anisotropies are a very long range, low energy phenomena and it seems 
unlikely to us that they can be cured by particle creation at very
negative times. 

As $t\ra -t_{string}$ copious particle production
and also other mechanisms, like  higher order corrections to the
action as conjectured in~\cite{giov} may damp anisotropies. But these 
corrections only become important close to
the Planck time. During the entire pre-big-bang inflation the
universe remains as anisotropic as in the initial
conditions. This has significant
implications on quantum particle creation during the  pre-big-bang
phase as has been investigated in~\cite{giov}.

Within string cosmology, cosmological fluctuations are due to
coherent quantum particle production in the pre-big-bang phase~\cite{DGSV}. It 
is still an open problem, to what extent anisotropic cosmological
fluctuations would be visible in the perturbations of the post-big-bang
universe, like {\it e.g.} as a preferred direction in the anisotropies
of the cosmic microwave background. A study of this problem is in
preparation.

{\it Acknowledgments} We would like to thank G. Veneziano
for discussions. This work is partially supported by the Swiss NSF.

\end{document}